\documentclass[fleqn,usenatbib]{mnras}

\usepackage{newtxtext,newtxmath}
\usepackage{anyfontsize}

\usepackage[T1]{fontenc}

\DeclareRobustCommand{\VAN}[3]{#2}
\let\VANthebibliography\thebibliography
\def\thebibliography{\DeclareRobustCommand{\VAN}[3]{##3}\VANthebibliography}

\usepackage{graphicx}	% Including figure files
\usepackage{amsmath}	% Advanced maths commands
\usepackage{subcaption}
\title[GNSS derived PWV at potential AMT sites]{A comparative analysis of GNSS-inferred precipitable water vapour at the potential sites for the Africa Millimetre Telescope}

\author[L. Frans, M. Backes, H. Falcke and T. Venturi]{
Lott Frans$^{1}$\thanks{E-mail: lfrans@unam.na},
Michael Backes$^{1,4}$,
Heino Falcke$^{2}$,
and Tiziana Venturi$^{3}$
\\
$^{1}$Department of Physics, Chemistry $\&$ Material Science, University of Namibia, Private Bag 13301, Windhoek, Namibia\\
$^{2}$Department of Astrophysics, Institute for Mathematics, Astrophysics and Particle Physics, Radboud University, P.O. Box 9010, 6500 GL Nijmegen, The Netherlands\\
$^{3}$Istituto di Radioastronomia, Instituto Nazionale di Astrofisica, Via Gobetti 101, 40129 Bologna, Italy\\
$^{4}$Centre for Space Research, North-West University, Private Bag X6001, Potchefstroom 2520, South Africa
}

\date{Accepted 2025 January 11. Received 2025 January 11; in original form 2024 October 31}

\pubyear{\the\year{}}

\begin{document}
\label{firstpage}
\pagerange{\pageref{firstpage}--\pageref{lastpage}}
\maketitle

\begin{abstract}
The Event Horizon Telescope~(EHT) is a network of antennas across the globe currently used to image super-massive black holes~(SMBHs) at a frequency of 230~GHz. Since the release of the image of M87$^\ast$ in 2019 and, subsequently, that of Sgr~A$^\ast$ in 2022 by the EHT collaboration, the focus has shifted to dynamically imaging SMBHs. This has led to a search for potential sites to extend and fill in the gaps within the EHT network. The Gamsberg Mountain and the H.E.S.S. site are both located within the Khomas highlands and have been identified as potential sites for the Africa Millimetre Telescope~(AMT). Precipitable water vapour~(PWV) in the atmosphere is the main source of opacity and noise from atmospheric emissions when observing at millimetre to sub-millimetre wavelengths. This study aims to establish the PWV content and the atmospheric transmission at 86, 230, and 345~GHz at the AMT potential sites using Global Navigation Satellite System~(GNSS) derived PWV data. Results show both sites have potential for observations at 86 and 230~GHz, with 345~GHz possible at the Gamsberg Mountain during winter. The overall median PWV of 14.27~mm and 9.25~mm was calculated at the H.E.S.S. site and the Gamsberg Mountain, respectively. The EHT window had PWV medians of 16.62~mm and 11.20~mm at the H.E.S.S. site and Gamsberg Mountain, respectively. Among the two sites, the Gamsberg Mountain had the lowest PWV conditions, therefore making it the most suitable site for the AMT.

\end{abstract}


%%%%%%%%%%%%%%%%%%%%%%%%%%%%%%%%%%%%%%%%%%%%%%%%%%

%%%%%%%%%%%%%%%%% BODY OF PAPER %%%%%%%%%%%%%%%%%%

\section{Introduction}
The Event Horizon Telescope~(EHT) is a network of antennas across the globe that operates at 230~GHz (1.3~mm) and is used in an interferometer~(VLBI) to image Super Massive Blackholes~(SMBHs). In 2017 the network with 7 antennas at 5 geographical sites and 8 antennas at 6 geographical sites across the globe observed M87* and Sgr~A*, respectively, between 5~April and 11~April~\citep{EHTM87,EHTSgrA}. The image of M87* was released in 2019~\citep{EHTM87} and that of Sgr~A* in 2022~\citep{EHTSgrA}. The release of these images prompted the next phase for the EHT, which is to dynamically image SMBHs and therefore make blackhole movies, further improve the angular resolution of the images, and add more robustness to the network. However, to make dynamical images, more antennas need to be added to the current configuration of the EHT. The Africa Millimetre telescope~(AMT) is one such antenna planned to be built in the Khomas highlands of Namibia~\citep{AMT_backes}.
The telescope is planned to be of a 15~m dish, either a new NOEMA-type dish~\citep{NOEMA_spec}, such as the one found on Plateau de~Bure in the French Alps, or the refurbished Swedish-ESO Submillimetre Telescope~(SEST)~\citep{sest_telescope} dish currently located in La~Silla, Chile~\citep{AMT_backes}. Two potential sites have been identified within the Khomas highlands for the AMT, these being the High Energy Stereoscopic System~(H.E.S.S.) observatory site and the Gamsberg Mountain, with the sites separated by a distance of 30~km. The H.E.S.S. observatory stands at 1859~m above sea level~(a.s.l.) and is a well-established very-high-energy gamma-ray observatory that consists of five imaging atmospheric Cherenkov telescopes~(IACTs)~\citep{hess_nam} whilst the Gamsberg Mountain is a flat-top mountain that stands at 2377~m~a.s.l.
Table~\ref{tab:site_info} summarizes the geographical information on the two sites as obtained by the Global Navigation Satellite System~(GNSS) stations.
\begin{table}
\caption{Geographical information on the studied sites as obtained by the GNSS stations.}
\label{tab:site_info}
	\centering
\begin{tabular}{c c c c}
    \hline
    Site & lat [deg] & long [deg] & altitude [m~asl]\\ [0.5ex]
    \hline%\hline
    H.E.S.S. site & -23.275 & 16.505 & 1859\\ 
    %\hline
    Gamsberg Mountain & -23.339 & 16.224 & 2377\\%[1ex] 
    \hline
\end{tabular}
\end{table}
Precipitable water vapour~(PWV) is the main source of opacity at millimetre to sub-millimetre wavelengths and is defined as the amount of water vapour in the atmospheric column above a location equivalent to the amount of liquid precipitation that would result if all the water vapour in the column was condensed~\citep{pwvdefinition}.\\
\\
On the Gamsberg Mountain, in-situ PWV measurements derived from sky emissivity at 18.5 microns have been obtained between 1994 and 1995 by \citet{eso_1994_5}. These measurements were taken every 2~hours during photometric nights. The results showed low PWV conditions with a yearly mean of 5.2~mm and yearly median of 5.0~mm. \citet{eso_1994_5} also found mean PWV values of 4.3~mm and below during the winter months of June, July, and August. During the EHT observing window period of March and April, a mean of 6.5~mm and 6.4~mm was found, respectively. Moreover, medians of 8.2~mm and 5.6~mm were observed during March and April, respectively.\\
\\
The H.E.S.S. site has numerous instruments onsite that are capable of measuring PWV and those from which PWV can be inferred, which include a NASA AErosol RObotic NETwork~(AERONET)~\citet{aeronet} station, the Autonomous Tool for Measuring Site COndition PrEcisely~(ATMOSCOPE)~\citep{Atmoscope} station, and infrared radiometers found on the H.E.S.S. Cherenkov Telescope~(CT)~1--4 telescopes herein referred to as CT-Meteo. A study by~\citet{fransisco_thesis} using these instruments at the H.E.S.S. site found the results from the different instruments not to agree due to their different biases. Data from the various instruments were not taken over the same period and the instruments also have different integration times~\citep{fransisco_thesis}. Moreover, some instruments such as the AERONET only take measurements during the daytime and IR radiometers on the H.E.S.S. telescope only take measurements during H.E.S.S. observations which occurs only during photometric nights~\citep{fransisco_thesis}. This resulted in the instruments having their biases which would reflect in the data, for example, the IR radiometers and the AERONET are likely to have low PWV results as they both conduct measurements only during optimal observing conditions. Furthermore, since H.E.S.S. observations only occur during photometric nights of which there is naturally more in winter than in summer, the IR radiometer data will be more likely skewed towards the winter period PWV data for which there are more H.E.S.S. observations than in summer. This will naturally result in low PWV values from the IR radiometers. The AERONET measurements are also only taken during cloudless daytime conditions, which would naturally occur more frequently during winter than in the summer period, which also skews it to low PWV values when compared to other instruments.
These biases are reflected in the results of~\citet{fransisco_thesis}, which shows a discrepancy in the data from the different sources. ~\citet{fransisco_thesis} found a clear seasonal trend in PWV across all the instruments with the PWV dropping to the lowest PWV values during the winter period and rising during the summer with the highest values. Overall,~\citet{fransisco_thesis} concluded the mean and median PWV values at the H.E.S.S. site to be 6.08~mm and 5.48~mm, respectively. Moreover, it was found that H.E.S.S. mean PWV rises to 15.79~mm in summer and drops to a low of 3.04~mm during winter~\citep{fransisco_thesis}.\\
\\
A study was conducted by~\citep{Potential_sites_Ray} to investigate candidate sites for the expansion of the EHT. In this study,~\citet{Potential_sites_Ray} used a 10-year MERRA-2 dataset interpolated over the location of each candidate site of 43 in total including the Gamsberg Mountain.~\citet{Potential_sites_Ray} reported an agreement between medians of opacity at 225~GHz calculated from MERRA-2 data and from a 225~GHz radiometer at EHT sites of Atacama Pathfinder EXperiment~(APEX), Large Millimeter Telescope~(LMT) and SubMillimeter Array~(SMA). Furthermore, a good agreement between MERRA-2-derived PWV values and field measurements was found for the driest and wettest sites within the EHT array. A median PWV value of less than 5~mm was reported by~\cite{Potential_sites_Ray} during March and April across the existing EHT sites. The Gamsberg Mountain was calculated to have a median opacity at 230~GHz of 0.84 and 0.56 during March and April, respectively~\cite{Potential_sites_Ray}.\\
\\
A study conducted by~\citet{GNSS_atacama} using GNSS station data in the Atacama desert to measure the PWV for sub-millimetre and millimetre observations found that GNSS station-derived PWV data could be reliably used for site evaluation and analysis. \citet{GNSS_atacama} compared the GNSS station-derived PWV data to scaled radiometer-derived PWV data and found that when the GNSS station was used with a barometer, it showed a systematic offset of -0.05~mm. Moreover,~\citet{GNSS_atacama} demonstrated that the statistical uncertainty of GNSS-derived PWV gets lower when averaging the PWV over a longer period by finding an uncertainty of 0.64~mm over 15~minutes, 0.52~mm over an hour, and 0.37~mm over 4~hours.\\
\\
In this study, GNSS station data from both the H.E.S.S. site and the Gamsberg Mountain, MERRA-2 data from both the H.E.S.S. site and Gamsberg Mountain were processed and converted into PWV for both sites. MERRA-2 data were used to find whether there is an agreement with the GNSS station data, and therefore validate and support the measurements from the GNSS station. Moreover, MERRA-2 data were used together with \texttt{am} to model and find the relationships between PWV and opacity at 86, 230, and 345~GHz for both sites. Furthermore, the PWV and atmospheric transmission at 86, 230, and 345~GHz was quantified during the EHT window of observations and the winter period which offered possibilities of single-dish and VLBI observations.

\section{Instruments and data acquisition}

\subsection{GNSS station}
\begin{figure}
    \centering
    \includegraphics[scale=0.1]{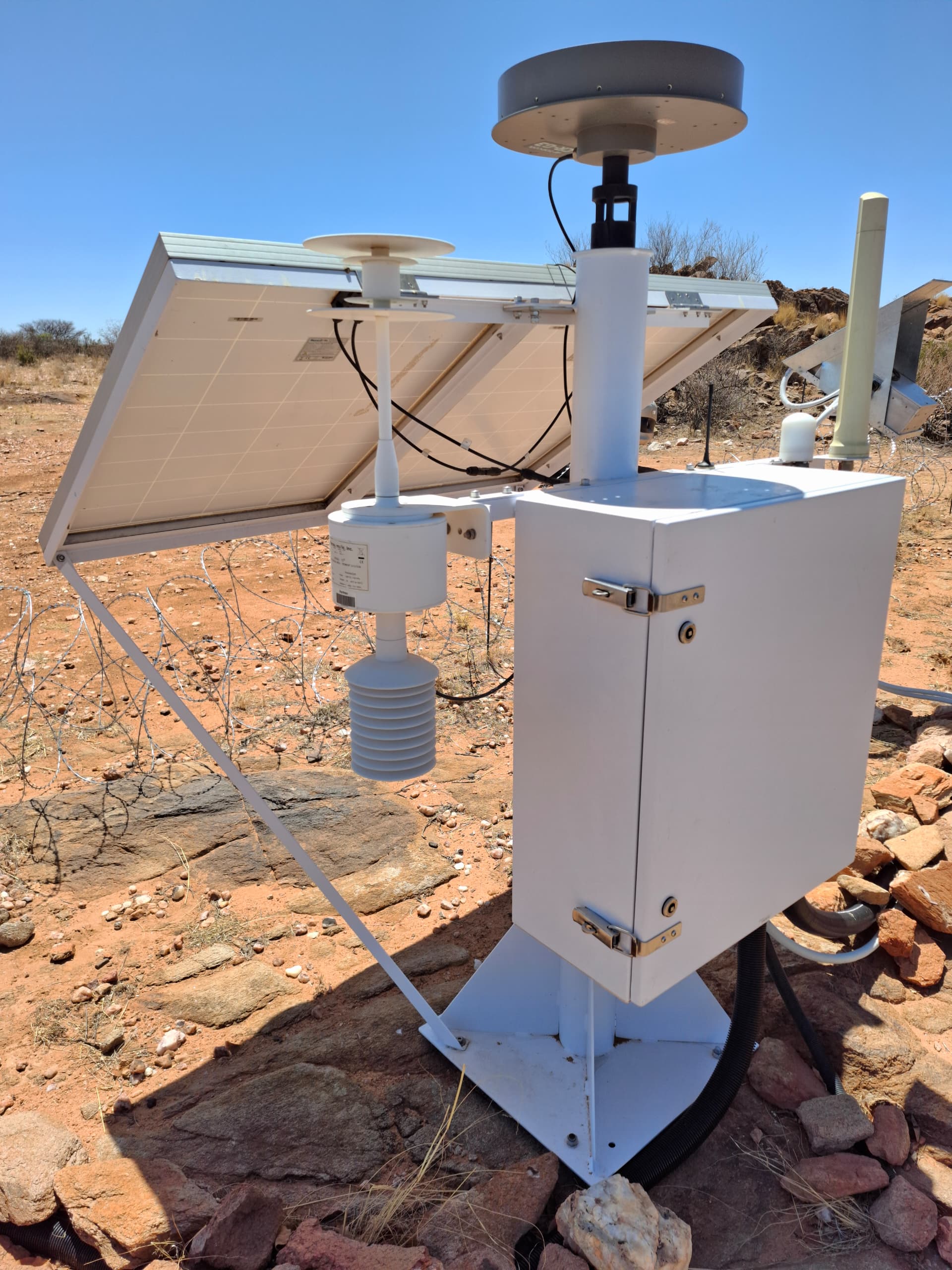}
    \caption{GNSS station with a MET4 weather station installed at the H.E.S.S. site. A similar GNSS station was installed at the Gamsberg Mountain.}
    \label{fig:gnss}
\end{figure}

A GNSS station was installed at both the H.E.S.S site and the Gamsberg Mountain. Figure~\ref{fig:gnss} shows the GNSS station at the H.E.S.S. site.
%A GNSS station was installed at both the H.E.S.S. site and the Gamsberg Mountain, figure~\ref{fig:gnss} shows the GNSS station at the H.E.S.S. site.
This station has been in operation and logging data since September 2022 whilst the Gamsberg Mountain station was installed in January 2024 and has been in operation since. The raw data measurements at both stations are integrated over 30-second intervals. The GNSS station can be used to directly estimate the integrated PWV using the measured zenith total delay $ZTD$, station latitude, station height, pressure, and temperature measured by the station. The $ZTD$ can be described in terms of equation~\ref{eq:ZTD},
\begin{equation}
ZTD = ZWD + ZHD
\label{eq:ZTD}
\end{equation}
where the $ZHD$ is the zenith hydrostatic delay and the $ZWD$ is the zenith wet delay~\citep{gnss_thesis_combrink,gnss_pwv_study_paper}. Using the meteorological parameters from \citet{saastamoinen1972atmospheric_model} and the \citet{davis1985_model} models, the $ZHD$ can then be calculated as,
\begin{equation}
ZHD = (2.2779\pm0.0024)\times\frac{P_s}{f(\lambda,H)}
\label{eq:ZHD}
\end{equation}
where $P_s$ is the total measured surface pressure at the GNSS station in hPa and $f(\lambda,H)$ is defined as,
\begin{equation}
f(\lambda,H) = 1 - 0.00266\times\cos({2\lambda})-0.00028H
\label{eq:function}
\end{equation}
where $H$ is the height in metres and~$\lambda$ the latitude of the GNSS station. Given the station measures the $ZTD$ and the $ZHD$ can be calculated using equation~\ref{eq:ZHD} and~\ref{eq:function}, then the $ZWD$ can be obtained from equation~\ref{eq:ZTD} as,
\begin{equation}
ZWD = ZTD - ZHD
\label{eq:ZWD}
\end{equation}
from which the integrated PWV~\citep{gnss_thesis_combrink} herein just referred to as PWV can be calculated in~mm as follows,
\begin{equation}
PWV = \frac{10^6}{\rho_w R_w \left( k^{'}_2 + k_3 T_{m}^{-1} \right) }\times ZWD
\label{eq:gnssPWV}
\end{equation}
where $\rho_w$ is the density of water given as 1000~kg\,m$^{-3}$, $R_w$ is the specific gas constant of water vapour, given as $461.4$ J\,K$^{-1}$\,kg$^{-1}$ and constants $k^{'}_2 = 22.1$~K\,hPa$^{-1}$ and $k_3 = 373900$~K$^2$\,hPa$^{-1}$ by ~\citet{Bevismodel}. $T_m$ is the weighted mean temperature of the column of water vapour above the GNSS station.\\
\\ 
The $ZTD$ GNSS data products used in this study from both stations were processed by the Nevada Geodetic Labratory~(NGL)~\citep{GNSS_nevada} using GipsyX version 1.0~\citep{gipsyx}. The NGL uses the Vienna mapping function~(VMF1) interpolated pressure ($P_{s}$) and mean atmospheric temperature $T_m$ in its processing of $ZHD$ and, eventually, PWV. However, for more accurate results, the pressure measured on site with high accuracy by the GNSS MET~4A weather station will be used in calculating $ZHD$ instead. For this, on-site pressure was integrated over from 30 seconds to 5 minutes in order to match the cadence of the NGL troposphere solution. The NGL products of $ZTD$ and $T_{m}$ were then time-matched to the on-site $P_{s}$ from which the PWV was calculated using the methods above.

\subsection{MERRA-2}
MERRA-2 assimilates into earth system modeling the upper air (radiosonde) measurements in addition to satellite and surface measurements. The dataset has a spatial resolution of $0.5^{\circ}$ latitude $\times 0.625^{\circ}$ longitude and 42 pressure levels~\citep{Merra2}. In this study, the dataset ranges over 24~years in length between 2000 and 2024 with a temporal resolution of 3~hours. MERRA-2 data were interpolated to what it would be at the H.E.S.S. site and the Gamsberg Mountain. Using \texttt{am} atmospheric model~\citep{am_atmospheric_model}, PWV, and opacity at 86, 230, and 345~GHz were then extracted from the interpolated MERRA-2 data for both the Gamsberg Mountain and the H.E.S.S. site.

\section{Methods and Results}
\begin{figure}
    \centering
    \includegraphics[scale=0.4]{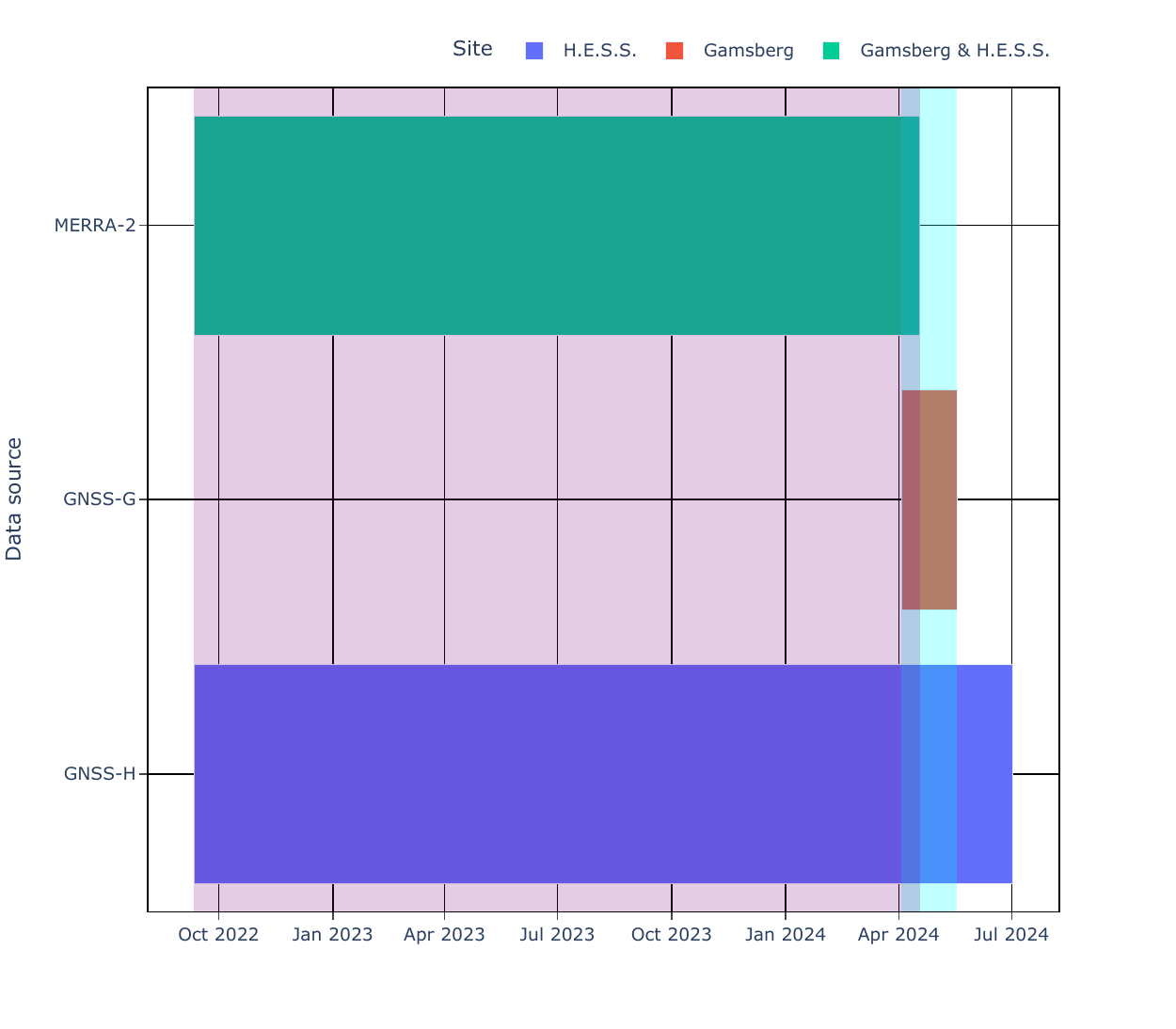}
    \caption{Periods for which data from the various sources are available for Gamsberg Mountain and the H.E.S.S. site. The MERRA-2 dataset in this study spans from 1 January 2000 to 18 April 2024 and is only displayed since September 2022.}
    \label{fig:instrument_plot}
\end{figure}

In this study, the focus will be on GNSS station data acquired from the H.E.S.S. site and the Gamsberg Mountain. MERRA-2 data will also be used to validate GNSS station data and to also acquire the relationships between PWV and opacity at 86, 230, and 345~GHz at both sites. Figure~\ref{fig:instrument_plot} summarizes the periods for which the data in this study is available for and the periods which the data overlaps. The different vertical colors give the period for which there is an overlap between the data from the sources. As can be seen from figure~\ref{fig:instrument_plot}, the data acquisition period of the GNSS station from the Gamsberg Mountain is short and only spans over 2 Months from 2 April 2024 and May 2024 even though the GNSS station was installed between 17 January. This is due to possible power outages of the GNSS station which causes the data not to be logged consistently and in some cases data to be lost. In order to get long-term GNSS station PWV data at Gamsberg Mountain, the differences between the Gamsberg Mountain and H.E.S.S. site PWV have to be found for periods of which there is consecutive GNSS station data at both sites. Thereafter the H.E.S.S. site PWV can be converted into Gamsberg Mountain PWV under the assumption the two sites experience the same conditions.

\subsection{H.E.S.S. site PWV}
\subsubsection{GNSS vs MERRA-2 validation}
\begin{figure*}
	\centering
	\begin{subfigure}[t]{1.\linewidth}
        \centering
        \includegraphics[scale=0.4]{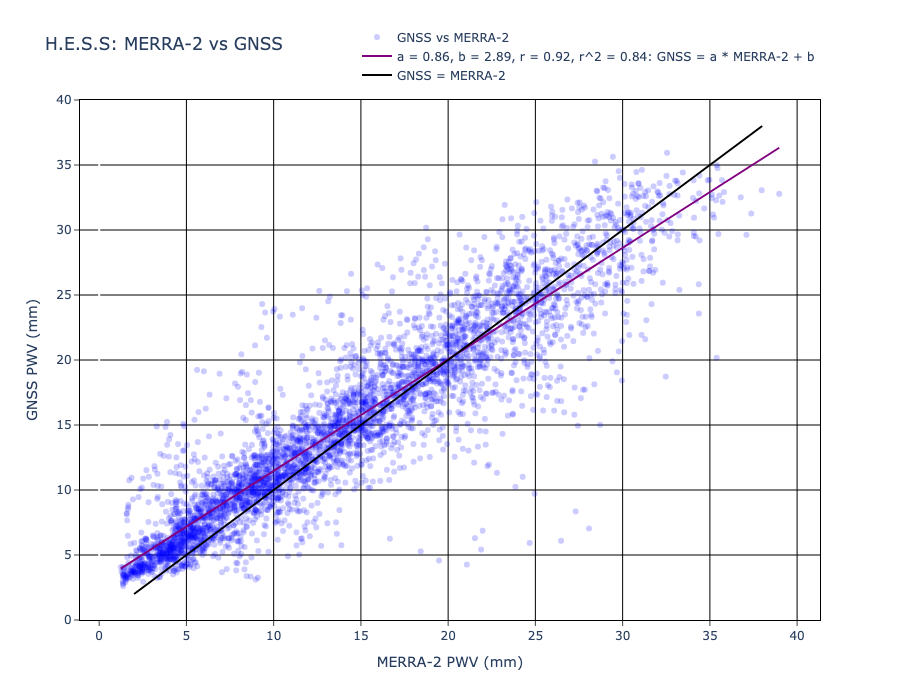}
		\caption{Correlation of MERRA-2 against GNSS station PWV measurements at the H.E.S.S. site.}
		\label{fig:merra_vs_gnss_corr}
	\end{subfigure}
	\begin{subfigure}[t]{1.\linewidth}
        \centering
		\includegraphics[scale= 0.16]{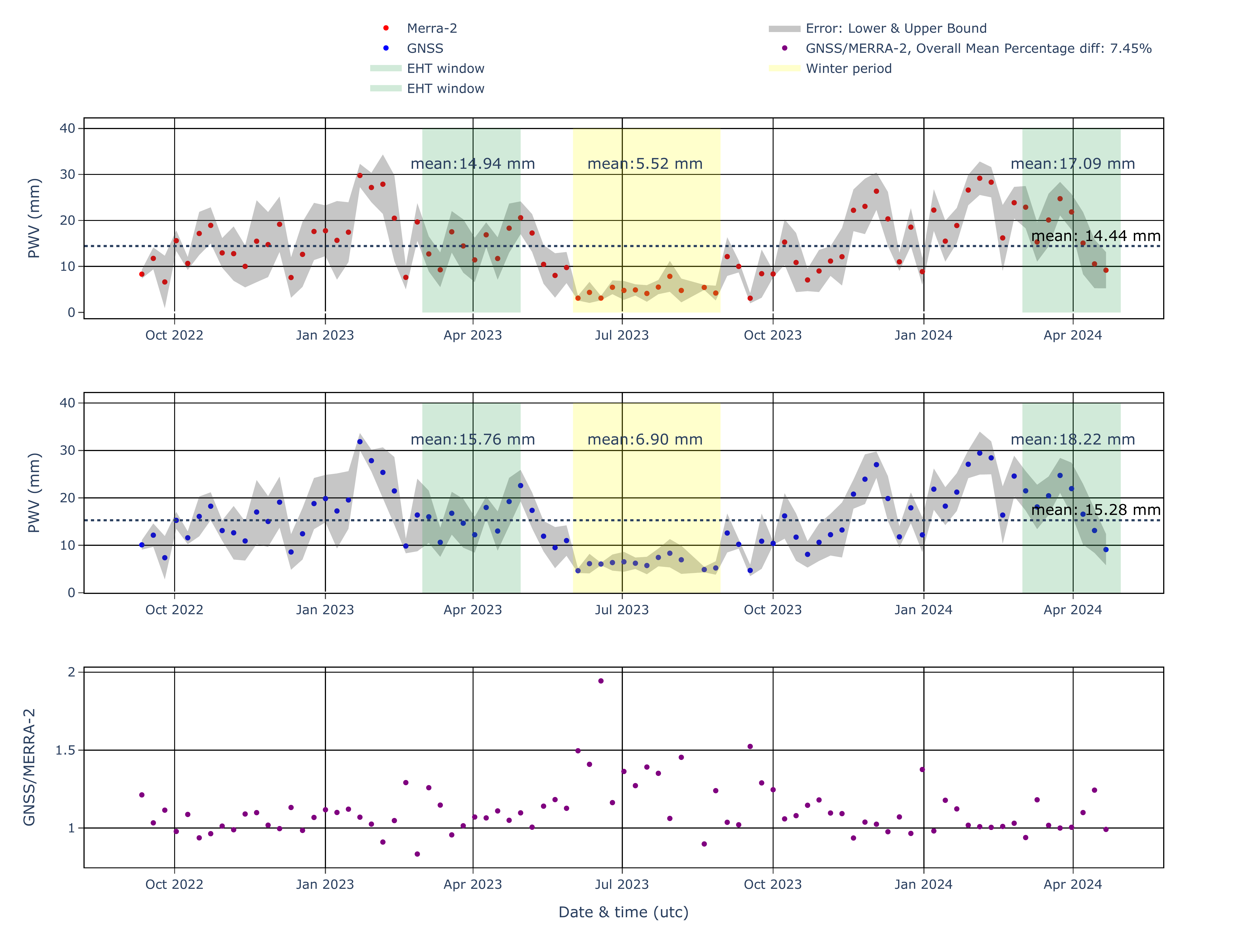}
		\caption{A comparison of the PWV data between the GNSS station and MERRA-2 PWV between September 2022 and April 2024. PWV during the EHT window of observing (shaded green) and winter period (yellow) was calculated.}
		\label{fig:gnss_vs_merra_monthly_data}
	\end{subfigure}
 \caption{Validation analysis of the PWV data from MERRA-2 and GNSS station.}
 \label{fig:merra_2_gnss_anal}
\end{figure*}

In order to analyze the differences in data between the two data sources, data taken over the same period between September 2022 and April 2024 (over a year) was analyzed. GNSS station data measurements were converted to a 3-hour temporal resolution to match MERRA-2's 3-hour temporal resolution. Figure~\ref{fig:merra_vs_gnss_corr} shows a plot of the data from the two sources plotted against each other.
Figure~\ref{fig:merra_vs_gnss_corr} shows a high positive correlation of 92\% between MERRA-2 and GNSS station data. A quick visual inspection of the trend in figure~\ref{fig:gnss_vs_merra_monthly_data} also shows that the two estimates follow a similar trend. However, each measured PWV value from the two estimates shows a percentage difference of 7.45\% on average. This results in a difference of 0.84~mm between MERRA-2's overall mean PWV of 14.44~mm and GNSS station's 15.28~mm during this period. Moreover, figure~\ref{fig:gnss_vs_merra_monthly_data} also shows the ratios of the estimated weekly PWV values to be centered around 1 but not exactly at 1. This shows the sources are measuring more or less the same PWV values over a week. A value of 1 would mean both instruments measured exactly the same PWV values for that specific week. The difference could be attributed to the mere fact that MERRA-2 PWV at the H.E.S.S. is obtained mainly from satellite data which are then estimated by interpolation for the H.E.S.S. site. Therefore, a radiometer needs to be installed in situ next to the GNSS station in order to validate the GNSS station data.

\subsubsection{H.E.S.S. site PWV against opacity modelling}
Since GNSS station and MERRA-2 PWV measurements are highly correlated, then MERRA-2 data can then be used to define the relationship between PWV and opacities at different frequencies. PWV and the equivalent opacity at different frequencies can be computed using \texttt{am}~\citep{am_atmospheric_model} from MERRA-2 data. Modelling the relationship between the PWV and opacity at any frequency will yield the model of PWV against opacity at that frequency for that specific site, in this case, the H.E.S.S. site.
\begin{figure*}
    \centering
    \includegraphics[scale=.4]{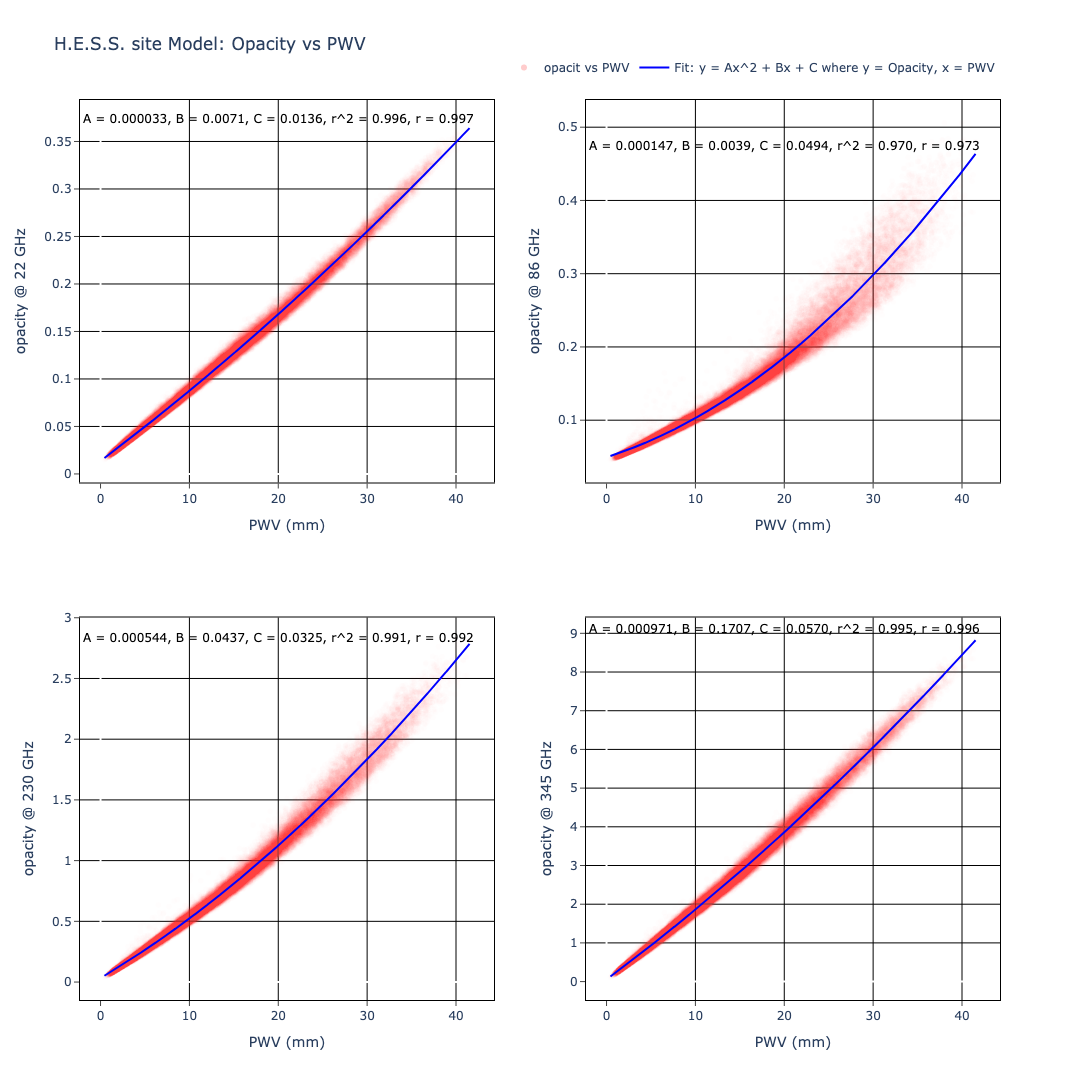}
    \caption{Fits of the H.E.S.S. site PWV against opacities at 22, 86, 230, and 345~GHz.}
    \label{fig:opacity_pwv_hess}
\end{figure*}
Figure~\ref{fig:opacity_pwv_hess} shows a polynomial fit of the relations between PWV in millimetres and opacity at frequencies of 22, 86, 230, and 345~GHz at the H.E.S.S. site. The latter 3 frequencies are the main frequencies of interest to this study. A polynomial of the form in equation~\ref{eq:poly} was fitted between PWV against all opacities at all frequencies, respectively.
\begin{equation}
\mathrm Opacity = A \times \mathrm PWV^2 + B \times \mathrm PWV + C
\label{eq:poly}
\end{equation}
The results of the fits between PWV and opacities at the different frequencies are listed in table~\ref{tab:PWV_opacity_hess}.
\begin{table}
\caption{Coefficients and coefficient of determination $r^2$ of the polynomial fit of equation~\ref{eq:poly} between PWV and opacity at 86, 230, and 345~GHz with their corresponding correlation coefficient $r$ at the H.E.S.S. site.}
\label{tab:PWV_opacity_hess}
	\centering
\begin{tabular}{c c c c c c}
    \hline
    Freq [GHz] & A~$\left[\textnormal{mm}^{-2}\right]$ & B~$\left[\textnormal{mm}^{-1}\right]$ & C & $r^2$ & $r$\\ [0.5ex] 
    \hline%\hline
    86 & 0.000147 &  0.0039 &  0.0494 & 0.970 & 0.973\\ 
    %\hline
    230 & 0.000544 & 0.0437 &  0.0325 & 0.991 & 0.992\\
    %\hline
    345 &  0.000971 &  0.1707 & 0.0570 & 0.995 & 0.996\\%[1ex] 
    \hline
\end{tabular}
\end{table}
As can be observed by the coefficient of determination, all the fits account for above 97\% of the variations in the data. These fits will then be used to convert the GNSS station PWV into opacity at the frequencies of interest at the H.E.S.S.

\subsubsection{H.E.S.S. site PWV, opacity, and atmospheric transmission}
\begin{figure*}
	\centering
	\begin{subfigure}[t]{1.\linewidth}
        \centering
		\includegraphics[scale=0.12]{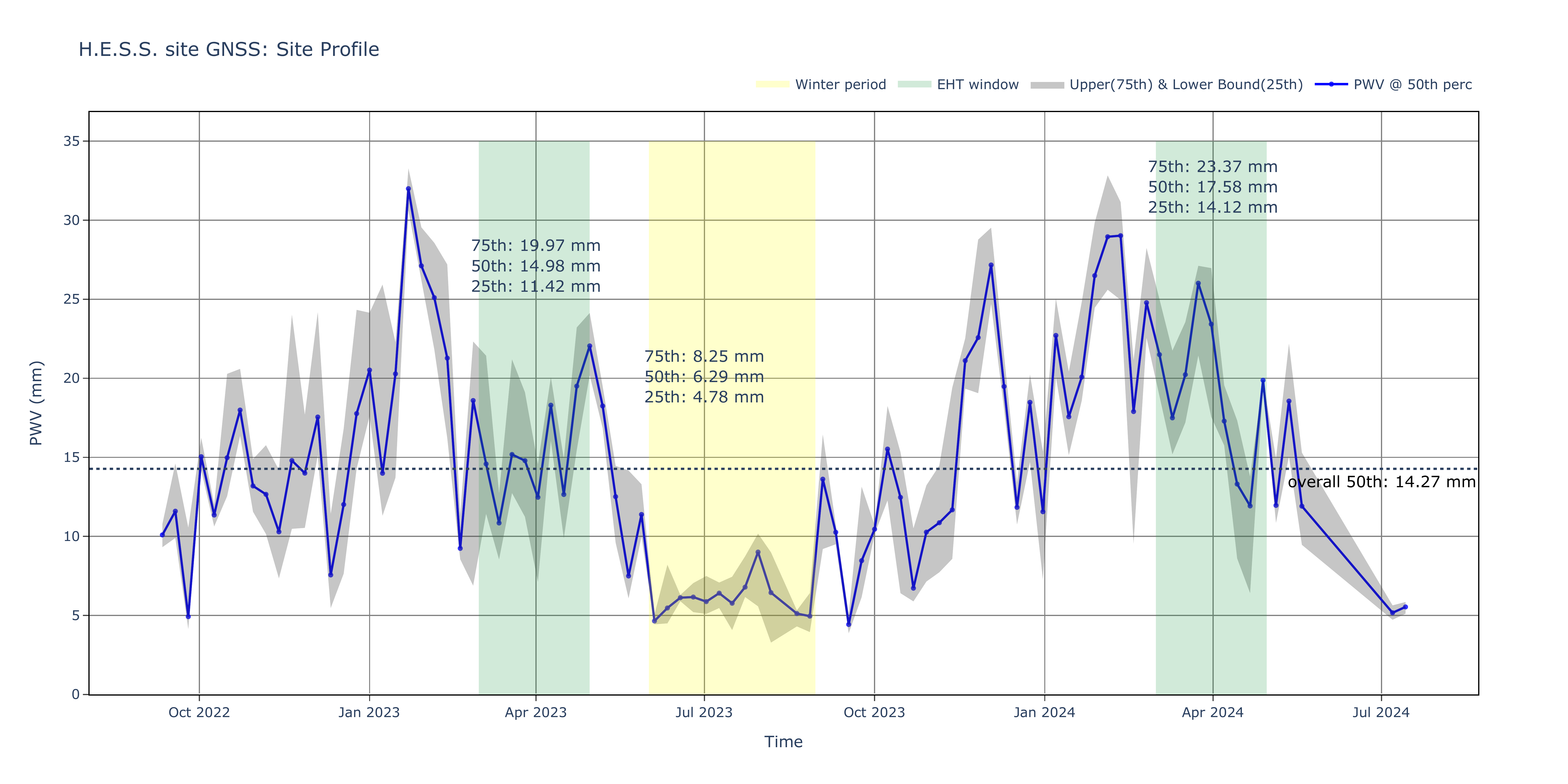}
		\caption{Weekly PWV as measured by the GNSS station at the H.E.S.S. site.}
		\label{fig:hess_trans}
	\end{subfigure}
	\begin{subfigure}[t]{1.\linewidth}
        \centering
		\includegraphics[scale=0.125]{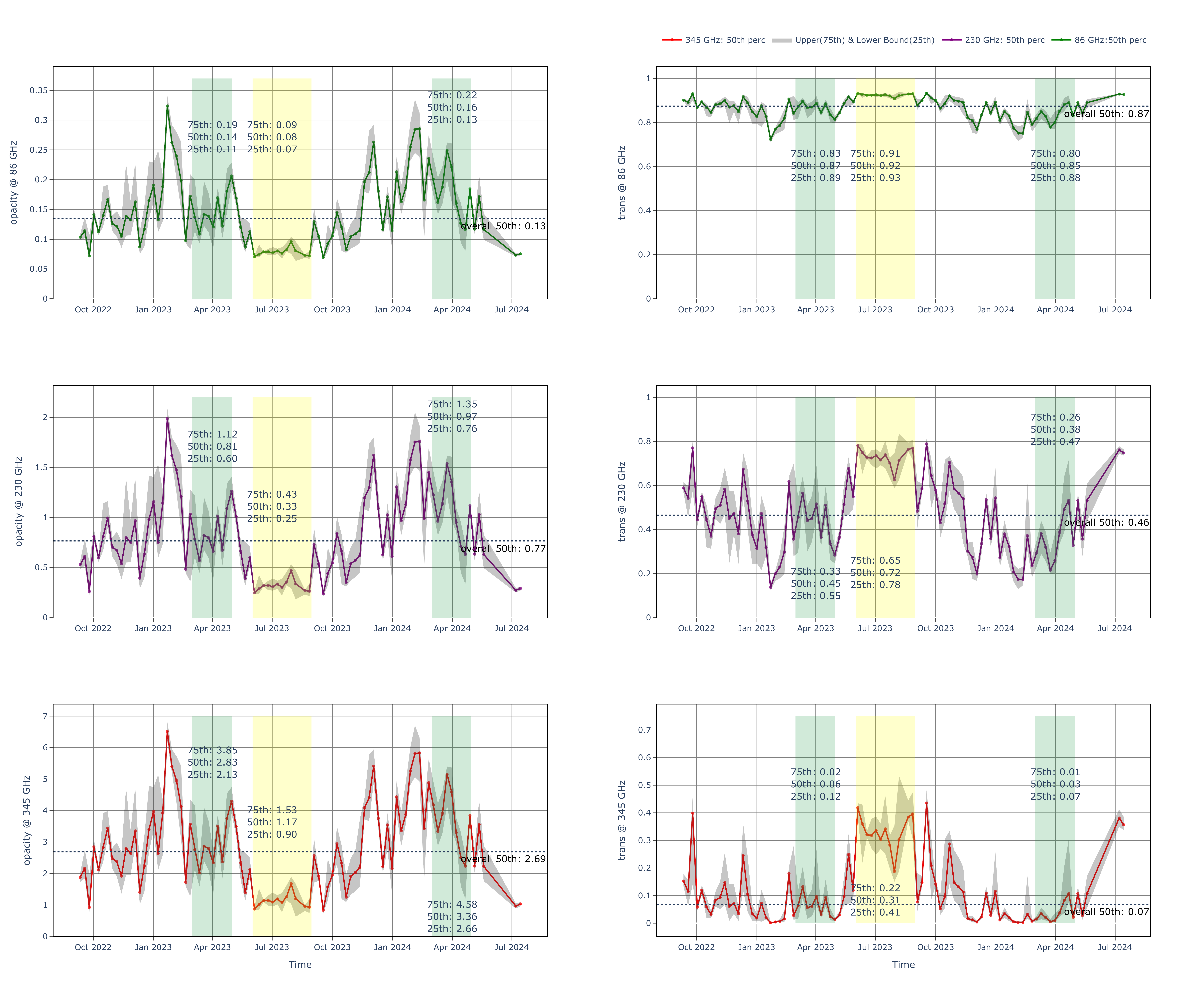}
		\caption{Weekly opacity as derived from PWV in figure~\ref{fig:hess_trans} using the fit relations from figure~\ref{fig:opacity_pwv_hess} and the corresponding atmospheric transmission at 86, 230, and 345~GHz.}
		\label{fig:hess_trans_summer}
	\end{subfigure}
 \caption{Weekly PWV, opacity, and atmospheric transmission at the H.E.S.S. site. The yellow period signifies the winter period of June, July, and August whilst the green is the EHT window of observations which occurs during March and April.}
 \label{fig:hess_trans_all}
\end{figure*}

Using the coefficients from table~\ref{tab:PWV_opacity_hess} in equation~\ref{eq:poly}, the PWV measured by the GNSS station was then used to calculate the opacities at 86, 230, and 345~GHz throughout the year. The atmospheric transmission at each frequency was then calculated for the same period using equation~\ref{eq:atransmission},
\begin{equation}
t(\nu) = e^{-\tau(\nu)}
\label{eq:atransmission}
\end{equation}
Where $\tau$ is the opacity~\cite{am_atmospheric_model}. A weekly 25$^{th}$, 50$^{th}$, and 75$^{th}$ percentiles of PWV, opacity, and atmospheric transmission were then calculated from the 5-minutes data sets. Figure~\ref{fig:hess_trans_all} shows the weekly results of PWV, opacity, and atmospheric transmission derived from GNSS station PWV.
As expected, the PWV and opacity are lowest across all frequencies during June, July, and August which is the southern hemisphere winter period. Conversely, the atmospheric transmission is highest across all frequencies during this period. The atmospheric transmission can also be observed to decrease across all frequencies during the summer period (December, January, and February) where the PWV is high. The overall 25$^{th}$, 50$^{th}$, 75$^{th}$ percentiles and mean results at each frequency is summarized in table~\ref{tab:hess_opacity}. The overall median PWV was calculated to be 14.27~mm, resulting in 87\% and 46\% atmospheric transmission at 86 and 230~GHz, respectively.
\begin{table}
\caption{The overall PWV with the corresponding opacity $\tau$ and atmospheric transmission $t$ at 86, 230, and 345~GHz at the H.E.S.S. site.}
\label{tab:hess_opacity}
	\centering
\begin{tabular}{c c c c c}
    \hline
    H.E.S.S. & 25\% & 50\% & 75\% & Mean\\ [0.5ex] 
    \hline
    PWV~[mm] & 9.24 & 14.27 & 20.47 & 15.20\\%[1ex]
     \hline
    $\tau_{86\text{~GHz}}$ & 0.10 & 0.13 & 0.19 & 0.15\\ 
    %\hline
    $\tau_{230\text{~GHz}}$ & 0.48 & 0.77 & 1.15 & 0.85\\
    %\hline 
    $\tau_{345\text{~GHz}}$ & 1.72 & 2.69 & 3.96 & 2.93\\%[1ex] 
    \hline
    $t_{86\text{~GHz}}$ & 0.91 & 0.87 & 0.83 & 0.86\\ 
    %\hline
    $t_{230\text{~GHz}}$ & 0.62 & 0.46 & 0.32 & 0.47\\
    %\hline
    $t_{345\text{~GHz}}$ & 0.18 & 0.07 & 0.02 & 0.12\\%[1ex] 
    \hline
\end{tabular}
\end{table}
Evidently, the best observing period for the AMT will be during the winter period. This period will be good for the AMT to conduct single dish observations at 86 and 230~GHz as at 75$^{th}$ percentile the atmosphere lets through 91\% and 65\% of the transmission at 86 and 230~GHz, respectively. At 345~GHz, with good conditions at 25$^{th}$ percentile, only 41\% of the transmission will be received by the AMT during winter. This period could also serve as an option for the AMT to conduct EHT observations from the H.E.S.S. site if the EHT were to conduct multiple observing campaigns beyond March and April in the future.

\subsubsection{Current EHT window at the H.E.S.S. site}
Only data at frequencies of 230 and 345~GHz were analyzed since the EHT primarily only observes at 230~GHz and has plans to expand to observing at 345~GHz in the future. It is evident from figure~\ref{fig:hess_trans} that the PWV for the EHT period of 2024 was higher than that of 2023. In order to get a good grasp of the atmospheric transmission during the EHT period, data from the two windows of the EHT campaigns of 2023 and 2024 were combined and analyzed with the results shown in table~\ref{tab:hess_opacity_eht}. Means and percentiles were calculated during this period. The total mean and median PWV across the two campaigns were found to be 16.61~mm and 16.62~mm. The atmospheric transmission at 230~GHz was found to have a median of 40\%, 31\% at the worst percentile of 75, and at the best percentile of 25, it is 52\%. The atmospheric transmission was found to be very low at 345~GHz with 10\% atmospheric transmission at 25 percentile.
\begin{table}
\caption{PWV and the corresponding opacity $\tau$ and atmospheric transmission $t$ at 230 and 345~GHz at the H.E.S.S. site during the EHT campaign period.}
\label{tab:hess_opacity_eht}
	\centering
\begin{tabular}{c c c c c}
    \hline
     H.E.S.S.$^\text{EHT}$& 25\% & 50\% & 75\% & Mean\\ [0.5ex] 
    \hline
    PWV~[mm] & 12.21 & 16.62 & 20.80 & 16.61\\%[1ex]
     \hline
    $\tau_{230\text{~GHz}}$ & 0.64 & 0.91 & 1.18 & 0.92\\
    %\hline
    $\tau_{345\text{~GHz}}$ & 2.29 & 3.16 & 4.03 & 3.20\\%[1ex] 
    \hline
    $t_{230\text{~GHz}}$ & 0.52 & 0.40 & 0.31 & 0.42\\
    %\hline
    $t_{345\text{~GHz}}$ & 0.10 & 0.04 & 0.01 & 0.08\\%[1ex] 
    \hline
\end{tabular}
\end{table}

\subsection{Gamsberg Mountain PWV}
\subsubsection{Gamsberg PWV against opacity modelling}
\begin{figure*}
    \centering
    \includegraphics[scale=0.4]{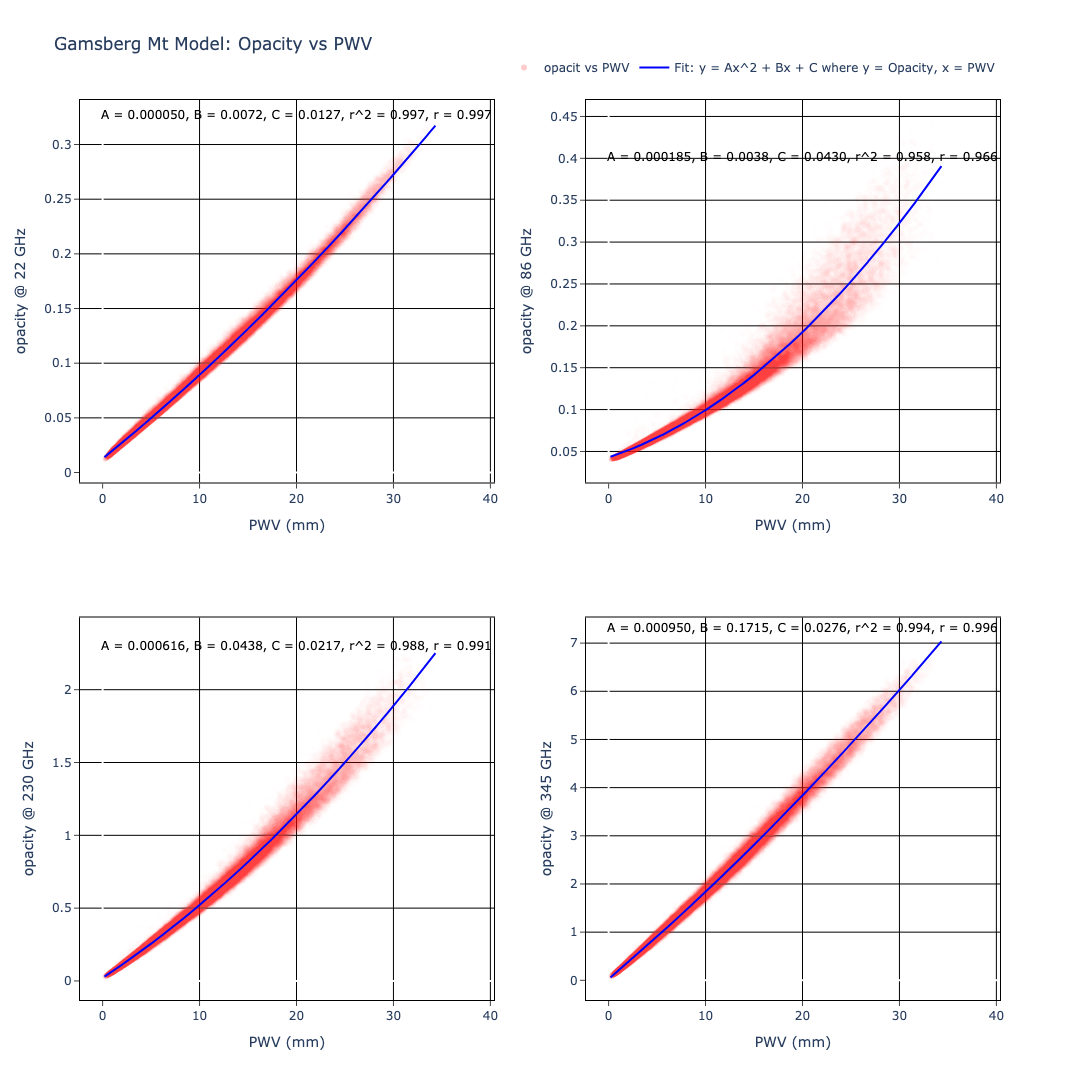}
    \caption{Fits of the Gamsberg Mountain PWV against opacity at 22, 86, 230, and 345~GHz.}
    \label{fig:opacity_pwv_gamsberg}
\end{figure*}

As was done for the H.E.S.S. site, the relationships between PWV and opacity at 86, 230, and 345~GHz were determined from MERRA-2 data using \texttt{am}~\citep{am_atmospheric_model} for the Gamsberg Mountain. As can be seen in figure~\ref{fig:opacity_pwv_gamsberg}, in all instances a polynomial of the form given by equation~\ref{eq:poly} was fitted to the data.
The coefficients and coefficient of determination of the fit are given in table~\ref{tab:PWV_opacity_gamsberg}. These relationships will be used to convert the GNSS derived PWV of the Gamsberg Mountain into opacity.
\begin{table}
\caption{Coefficients and coefficient of determination $r^2$ of the polynomial fit of equation~\ref{eq:poly} between PWV and opacity at 86, 230, and 345~GHz with their corresponding correlation coefficient $r$ at the Gamsberg Mountain.}
\label{tab:PWV_opacity_gamsberg}
	\centering
\begin{tabular}{c c c c c c}
    \hline
    Freq [GHz] & A~$\left[\textnormal{mm}^{-2}\right]$ & B~$\left[\textnormal{mm}^{-1}\right]$ & C & $r^2$ & $r$\\ [0.5ex] 
    \hline%\hline
    86 & 0.000185 & 0.0038 & 0.0430 & 0.958 & 0.966\\ 
    %\hline
    230 & 0.000616 & 0.0438 & 0.0217 & 0.988 & 0.991\\
    %\hline
    345 & 0.000950 & 0.1715 & 0.0276 & 0.994 & 0.996\\%[1ex] 
    \hline
\end{tabular}
\end{table}

\subsubsection{Gamsberg Mountain PWV from H.E.S.S. site PWV}
\begin{figure*}
    \centering
    \includegraphics[scale=0.15]{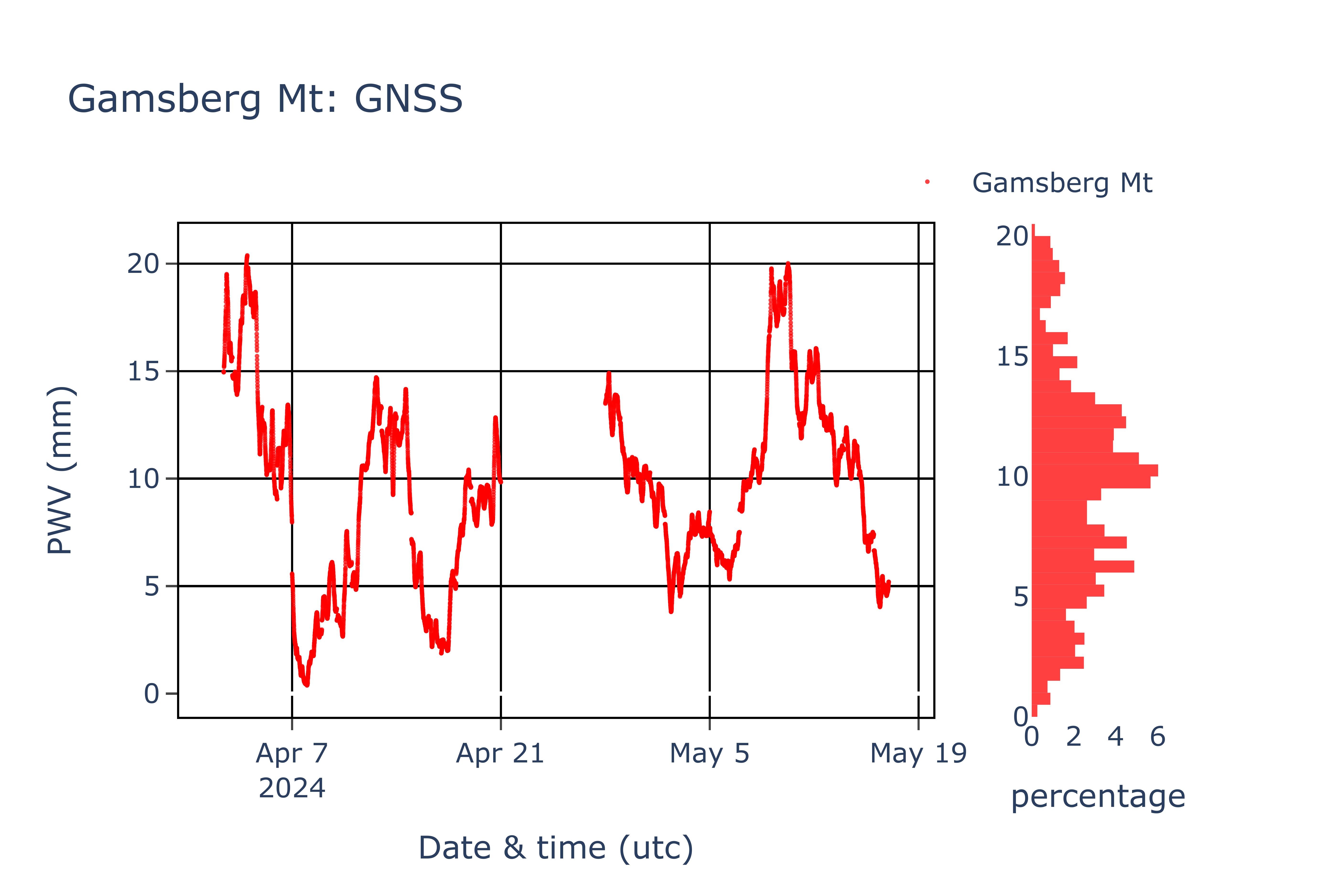}
    \caption{Data measured by GNSS station at the Gamsberg Mountain since 2 April.}
    \label{fig:gams_nevada_raw}
\end{figure*}

\begin{figure*}
    %\centering
    %\includegraphics[scale=0.16]{plots/hess_gams_site_comp2_neveda.pdf}
    \includegraphics[scale=0.17]{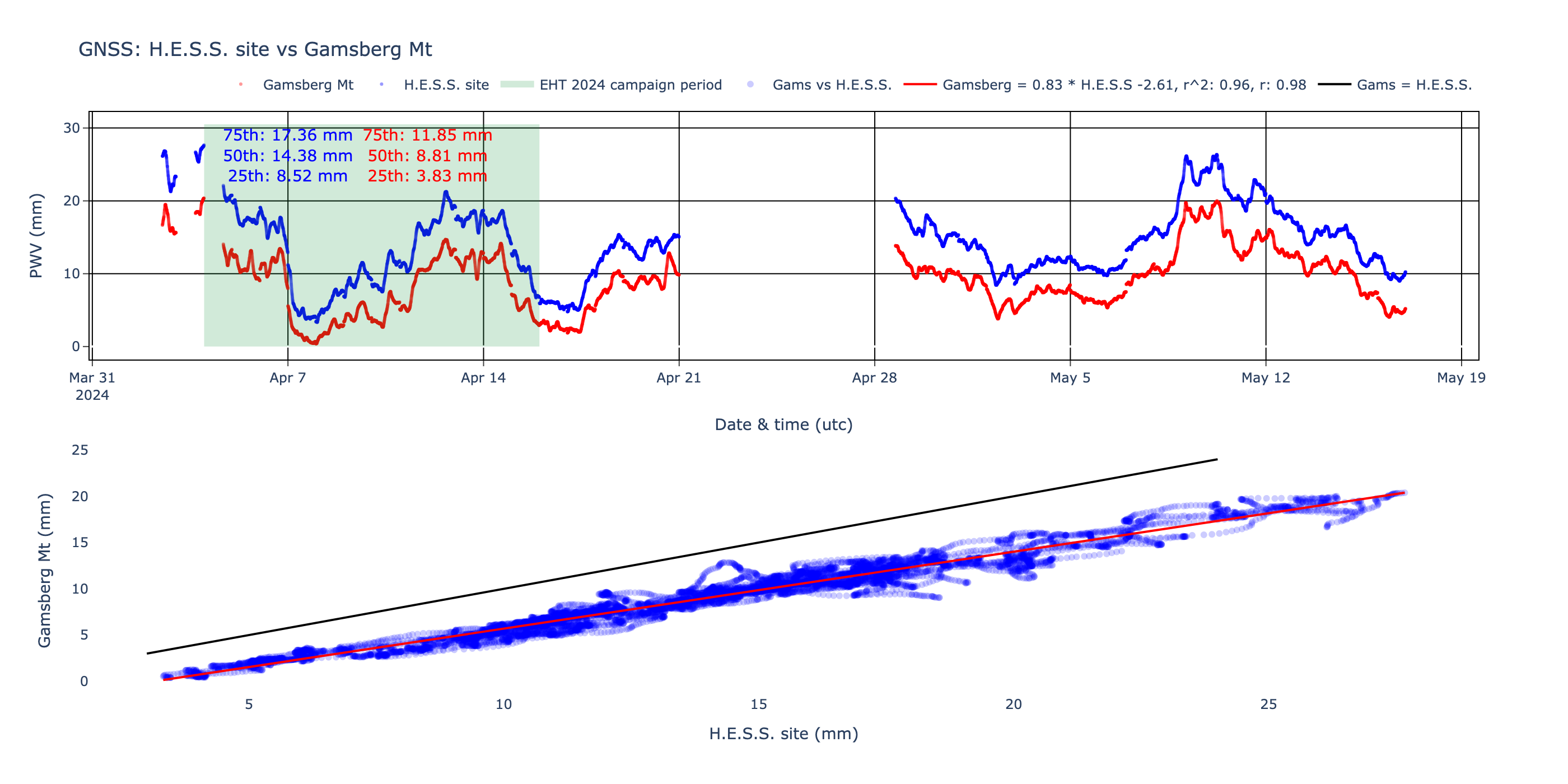}
    \caption{Data taken consecutively by GNSS stations at the Gamsberg Mountain and the H.E.S.S. site since 2~April 2024. %Blue and grey shows data taken at the H.E.S.S. site whilst red shows data taken at the Mt.~Gamsberg site. Blue and red data points indicate data taken contemporaneously on both sides.
    The lower figure shows the relation of the fit between Gamsberg Mountain and the H.E.S.S. site PWV.}
    \label{fig:gams_hess_compare}
\end{figure*}

The PWV at Gamsberg Mountain was estimated from the H.E.S.S. site PWV since the GNSS station at Gamsberg Mountain was set up approximately a year later than that of the H.E.S.S. site. Moreover, the Gamsberg GNSS station had only recorded data for a short period from April~2, 2024, to May~16, 2024 at the writing of this publication as can be seen in figure~\ref{fig:gams_nevada_raw}. More importantly, this was the period for which there were consecutive data from the GNSS station installed at both sites.
This would then allow to determine the relationship that equates the differences between the GNSS at the two sites and therefore allow for the conversion of the data from the H.E.S.S. site GNSS station into that of the Gamsberg Mountain with the assumption both sites experience the same atmospheric conditions as they are in the same locality. Figure~\ref{fig:gams_hess_compare} shows PWV data measured at the same time ever since the stations were consecutively running at both sites.
A linear relationship accounting for the differences in PWV between the two sites was observed as can be seen in figure~\ref{fig:gams_hess_compare}. The PWV from the two sites had a positive correlation of 98\%. The PWV from both sites are then linearly related by,
\begin{equation}
PWV_\text{Gam} = 0.83 * PWV_\text{H.E.S.S.} - 2.61 \textnormal{mm}
\label{eq:PWV_gam_vs_hess}
\end{equation}
where this relationship will account for $r^2 = 0.96$ of the data.

\subsubsection{Gamsberg Mountain PWV, opacity, and atmospheric transmission}
\begin{figure*}
	\centering
	\begin{subfigure}[t]{1.\linewidth}
        \centering
		\includegraphics[scale=0.12]{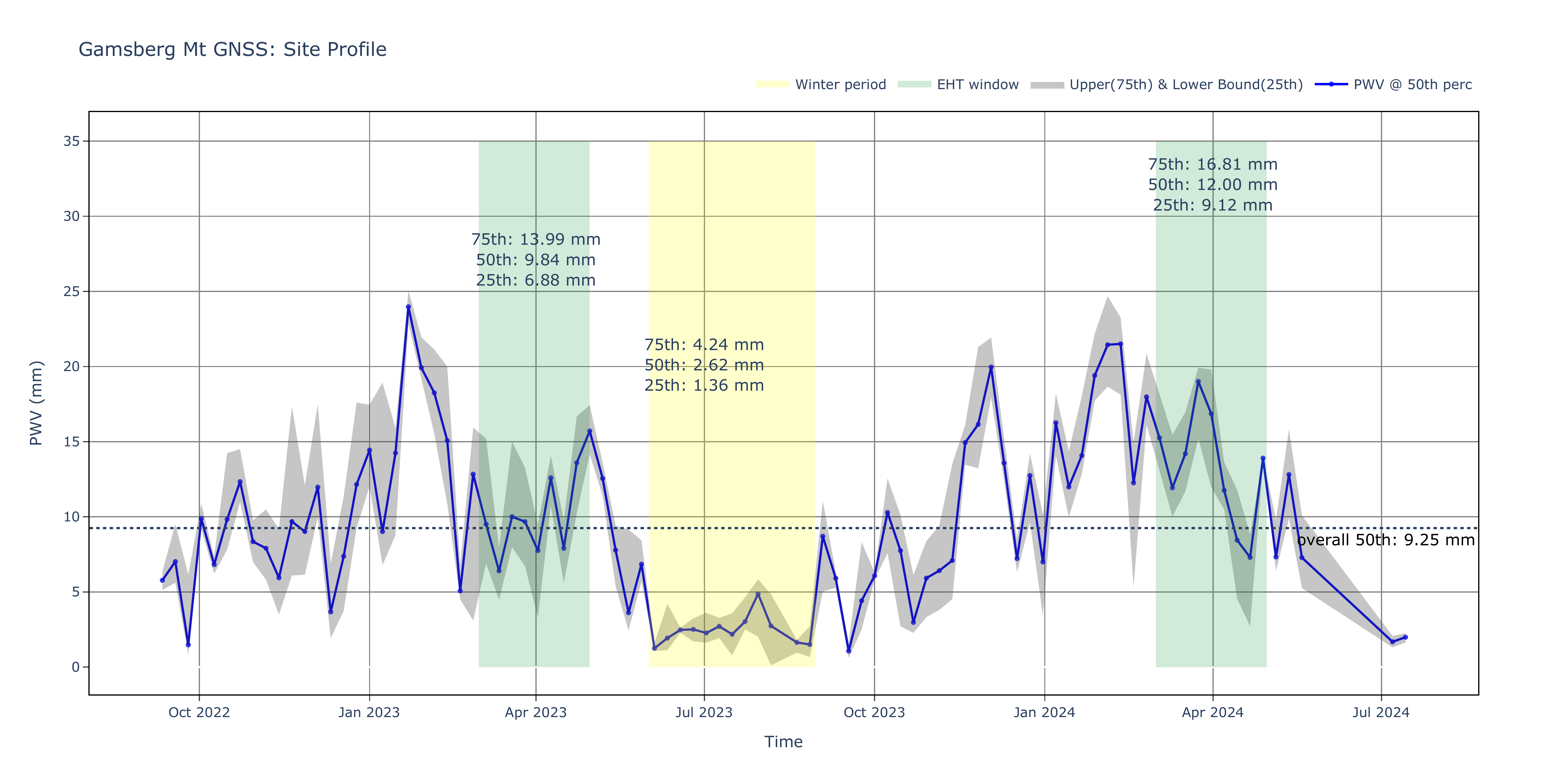}
		\caption{Weekly PWV at Gamsberg Mountain as calculated from H.E.S.S. site PWV data using equation~\ref{eq:PWV_gam_vs_hess}.}
		\label{fig:gamsberg_trans_1}
	\end{subfigure}
	\begin{subfigure}[t]{1.\linewidth}
        \centering
		\includegraphics[scale=0.125]{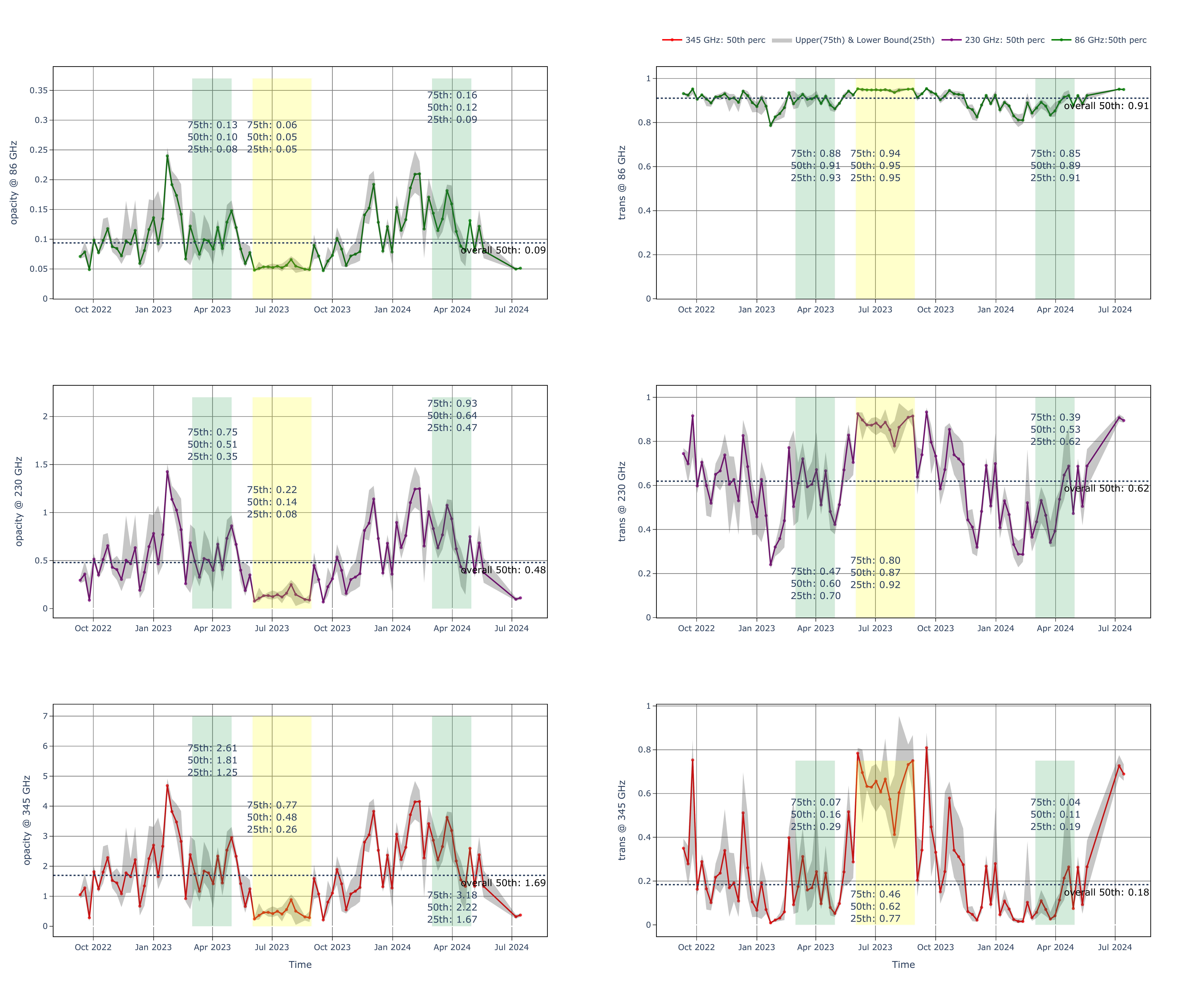}
		\caption{Weekly opacity as derived from PWV in figure~\ref{fig:gamsberg_trans_1} using the fit relations from figure~\ref{fig:opacity_pwv_gamsberg} and the corresponding atmospheric transmission at 86, 230, and 345~GHz.}
		\label{fig:gamsberg_trans_2}
	\end{subfigure}
 \caption{Weekly PWV, opacity, and atmospheric transmission at the Gamsberg Mountain. The yellow period signifies the winter period of June, July, and August whilst the green is the EHT window of observations which occurs during March and April.}
 \label{fig:gamsberg_all_trans}
\end{figure*}

Using equation~\ref{eq:PWV_gam_vs_hess}, and the assumption that the Gamsberg Mountain experiences the same atmospheric conditions as the H.E.S.S. site, the H.E.S.S. site PWV was converted into Gamsberg Mt PWV. Using the coefficient in table~\ref{tab:PWV_opacity_gamsberg} with equation~\ref{eq:poly}, the PWV was converted into opacity at 86, 230, and 345~GHz. Using equation~\ref{eq:atransmission}, the opacity at 86, 230, and 345~GHz was converted into the atmospheric transmission at 86, 230, and 345~GHz. Figure~\ref{fig:gamsberg_all_trans} shows the PWV, opacity, and atmospheric transmission at 86, 230, and 345~GHz of the Gamsberg Mountain estimated from the H.E.S.S. site PWV.
The overall results of the Gamsberg Mountain results are tabulated in table~\ref{tab:gams_opacity}.
\begin{table}
\caption{The overall PWV with the corresponding opacity $\tau$ and atmospheric transmission $t$ at 86, 230, and 345~GHz at the Gamsberg Mountain.}
\label{tab:gams_opacity}
	\centering
\begin{tabular}{c c c c c}
 \hline
  Mt.~Gamsberg& 25\% & 50\% & 75\% & Mean\\ [0.5ex] 
 \hline
    PWV~[mm] & 5.07 &  9.25 & 14.40 & 10.02\\%[1ex]
\hline
 $\tau_{86\text{~GHz}}$ & 0.07 & 0.09 & 0.14 & 0.11\\ 
 %\hline
 $\tau_{230\text{~GHz}}$ & 0.26 & 0.48 & 0.78 & 0.55\\
 %\hline
 $\tau_{345\text{~GHz}}$ & 0.92 & 1.69 & 2.69 & 1.88\\%[1ex] 
 \hline
 $t_{86\text{~GHz}}$ & 0.94 & 0.91 & 0.87 & 0.90\\ 
 %\hline
 $t_{230\text{~GHz}}$ & 0.77 & 0.62 & 0.46 & 0.62\\
 %\hline
 $t_{345\text{~GHz}}$ & 0.40 & 0.18 & 0.07 & 0.27\\%[1ex] 
 \hline
\end{tabular}
\end{table}
As expected, the seasonal variations are similar to that of the H.E.S.S. site in which the atmospheric transmission is highest in the southern hemisphere winter period and lowest during the summer as can be seen in figure~\ref{fig:gamsberg_trans_1}. At 86 and 230~GHz, the overall atmospheric transmission was very high with a median of 91\% and 62\%, respectively, whilst amounting to 18\% at 345~GHz. The best periods with the highest atmospheric transmission at all frequencies of interest within the year are during the southern hemisphere winter season during the period of June, July, and August. During this period, at 25$^{th}$ percentile, the PWV was found to be 4.24~mm. At the worst of 75$^{th}$ percentile, the atmospheric transmission at 86 and 230~GHz was 94\% and 80\%, respectively. At 345~GHz, the median was found to be 62\% and at 25$^{th}$ percentile when the conditions are favourable the atmosphere can let through 77\% of the transmission. This means that the AMT could observe at 86 and 230~GHz across all seasons throughout most of the year with the additional frequency at 345~GHz also only possible during the winter period (June, July, and August) from the Gamsberg Mountain.

\subsubsection{Current EHT window Gamsberg Mountain}
As was done for the H.E.S.S. site, the overall PWV, opacity, and transmission at the desired frequencies of 230 and 345~GHz were evaluated across the two EHT windows (March and April) in 2023 and 2024 as given in table~\ref{tab:gams_opacity_eht}. The PWV across the two campaigns resulted in a PWV median of 11.20~mm and a mean of 11.19~mm. The median atmospheric transmission at 230~GHz was found to be 55\% and at the worst 75 percentile was 45\%. At 345~GHz, the atmospheric transmission was very low with a median of 13\% and at the best 15\%.
\begin{table}
\caption{PWV and the corresponding opacity $\tau$ and atmospheric transmission $t$ at 230 and 345~GHz at the Gamsberg Mountain during the EHT campaign period.}
\label{tab:gams_opacity_eht}
	\centering
\begin{tabular}{c c c c c}
 \hline
  Mt.~Gamsberg$^\text{EHT}$& 25\% & 50\% & 75\% & Mean\\ [0.5ex] 
 \hline
   PWV~[mm]  & 7.54 &  11.20 & 14.67&  11.19\\%[1ex]
\hline
 $\tau_{230\text{~GHz}}$ & 0.39 & 0.59 &  0.80 & 0.60\\
 $\tau_{345\text{~GHz}}$ & 1.37 & 2.07 & 2.75 & 2.08\\%[1ex] 
 \hline%\hline
 $t_{230\text{~GHz}}$ & 0.68 & 0.55 & 0.45 & 0.57\\
 $t_{345\text{~GHz}}$ & 0.25 & 0.13 & 0.06 & 0.19\\%[1ex] 
 \hline
\end{tabular}
\end{table}
The AMT will have the potential to partake in EHT observations at 230~GHz during the current window of March and April from the Gamsberg Mountain. However, the atmospheric transmission at 345~GHz is very low making it difficult for observations during this period.

\section{Conclusions}
In this study, GNSS station data were validated against MERRA-2 data with measurements from both instruments found to be in agreement with a 92\% correlation. However, on average a percentage difference of 7.45\% was found between each PWV measurement of the GNSS station and MERRA-2. The difference in PWV measurements between MERRA-2 and GNSS station data can be attributed to the fact that MERRA-2 data are not exactly measured by an instrument at the H.E.S.S. site or the Gamsberg Mountain but are rather estimated through interpolation of satellite data and therefore varying from GNSS station measurements. In order to conduct a proper validation of GNSS PWV data, it is recommended a radiometer be installed alongside the GNSS station at the H.E.S.S. site. We have also shown in this study the relation between the H.E.S.S. site PWV and Gamsberg Mountain PWV data with the relation having an accuracy of 96\%.\\
\\
Overall the H.E.S.S. site median and mean PWV were found to be 14.27~mm and 15.20~mm, respectively, whilst the median and mean at the Gamsberg Mountain were found to be 9.25~mm and 10.02~mm. Whilst these Gamsberg Mountain values are much higher when compared to the median and mean of 5.0~mm and 5.2~mm, respectively measured at the Gamsberg Mountain in 1994/1995~\citep{eso_1994_5}, it should be noted that the latter study only considered photometric nights which skews the results towards lower PWV values.\\
\\
Overall, the H.E.S.S. site had a median atmospheric transmission at 86 and 230~GHz of 87\% and 46\%, respectively, whilst the Gamsberg Mountain had an overall median atmospheric transmission at 86 and 230~GHz of 91\% and 62\%, respectively. The seasonal trend in the region is such that during the southern hemisphere summer months which occurs during January, February, and December there is high PWV which drastically drops to low values during the winter months of June, July, and August. Both the Gamsberg Mountain and the H.E.S.S. site data show high values during the EHT window of observations. Over the two EHT campaigns of 2023 and 2024, the PWV was found at the H.E.S.S. site to have a median of 16.62~mm, whilst the Gamsberg Mountain recorded a median of 11.20~mm. If the AMT were to be built at the H.E.S.S. site it would only be able to partake in the EHT observations during the EHT window at 230~GHz in which the median atmospheric transmission is 40\% and when the conditions are favourable at 25$^{th}$ percentile, the atmospheric transmission becomes 52\%. At 345~GHz, observations may not be possible from the H.E.S.S. site as during this period at best 25$^{th}$ percentile, only 10\% of the signal will be transmitted through the atmosphere. In contrast, the AMT at the Gamsberg Mountain will receive a median of 55\% and at 25$^{th}$ percentile, 68\% of the emission at 230~GHz will be transmitted through the atmosphere making it more reliable to partake in EHT observations during this period. However, just as at the H.E.S.S. site, the atmospheric transmission at 345~GHz is rather low as at a best of 25$^{th}$ percentile only 25\% of the transmission make it to the ground during this period. The winter months period offers the best PWV conditions for observations at both sites. The highest atmospheric transmission throughout the year is observed across all frequencies during this period. During this period the median PWV drops to 6.29~mm and 2.62~mm for the H.E.S.S. site and the Gamsberg Mountain, respectively. The atmospheric transmission during winter at the Gamsberg Mountain at 345~GHz, had a median of 62\% and 77\% at 25$^{th}$ percentile when the conditions are favourable.\\
\\
In conclusion, whilst building the AMT at the H.E.S.S. site will save up on costs incurred on the construction of the observatory, it is evident from the data that the best site with the lowest PWV among the two sites is Gamsberg Mountain. This is expected as the mountain has a height advantage of 518~m over the H.E.S.S. site. Both the Gamsberg Mountain and the H.E.S.S. site have potential for observations at 86 and 230~GHz, whilst the Gamsberg has potential for observations at 345~GHz during the winter (June, July, and August) period. Generally, the EHT period has high PWV and low atmospheric transmission at 230 and 345~GHz with the highest potential for observations during this period occurring at Gamsberg Mountain at 230~GHz. With the EHT exploring options to conduct more observations outside of the current EHT window of March and April, the winter period will serve as one of the best periods to have the AMT join observations at 230~GHz from both sites. Moreover, If the AMT is to be built on Gamsberg Mountain, there will be potential to observe at 345~GHz during winter especially using the frequency phase transfer technique to improve the signal recovery. The outlook from this study is to install a radiometer in situ at the H.E.S.S. site and validate the GNSS data used in this study with on-site radiometer data.  

\section*{Acknowledgements}

We greatly appreciate the support by the South African Radio Astronomy observatory~(SARAO) through Roelf Botha and his Geodesy team for the installation and maintenance of GNSS stations at the Gamsberg Mountain and at the H.E.S.S. site. Furthermore, we also thank the Nevada Geodetic Laboratory~(NGL) for processing and presenting the GNSS tropospheric products used in this study. This work has been partially supported by the ERC Synergy Grant BlackHolistic and H2020-INFRADEV-2016-1 project 730884 JUMPING-JIVE.

%%%%%%%%%%%%%%%%%%%%%%%%%%%%%%%%%%%%%%%%%%%%%%%%%%
\section*{Data Availability}
The GNSS and MERRA-2 data set supporting this study is available from the authors upon reasonable request. Alternatively, Processed GNSS station data for the H.E.S.S. site station~(GBGA) can be accessed here (\url{http://geodesy.unr.edu/NGLStationPages/stations/GBGA.sta}) and for the Gamsberg Mountain station~(GBGB) here \url{http://geodesy.unr.edu/NGLStationPages/stations/GBGB.sta}).



\bibliographystyle{mnras}
\bibliography{references} % if your bibtex file is called example.bib

%\begin{thebibliography}{99}
%\end{thebibliography}

%%%%%%%%%%%%%%%%%%%%%%%%%%%%%%%%%%%%%%%%%%%%%%%%%%

%%%%%%%%%%%%%%%%% APPENDICES %%%%%%%%%%%%%%%%%%%%%

%\appendix



%%%%%%%%%%%%%%%%%%%%%%%%%%%%%%%%%%%%%%%%%%%%%%%%%%

% Don't change these lines
\bsp	% typesetting comment
\label{lastpage}
\end{document}